\documentclass[preprint2]{aastex}
\def\RSUN{\rm R$_{\odot}$}

\def\RSUN{\rm R$_{\odot}$}

\shorttitle{UV Spectra of Fast CMEs}
\shortauthors{Raymond et al.}

\begin{document}
\title{Far UV Spectra of Fast Coronal Mass Ejections Associated with X-class Flares}
\author{
J.C. Raymond, A. Ciaravella, D. Dobrzycka\altaffilmark{1}, L. Strachan, Y.-K. Ko and M. Uzzo}
\affil{Harvard-Smithsonian Center for Astrophysics, 60 Garden St., Cambridge, MA 02138}

\and

\author{Nour-Eddine Raouafi}
\affil{Max Planck Institut f\"{u}r Aeronomie, Katlenburg-Lindau, D 37191, Germany}

\altaffiltext{1}{also European Southern Observatory, Karl-Schwarzschild Str. 2, Garching, D-85748, Germany}

%
%


\begin{abstract}
The Ultraviolet Coronagraph Spectrometer (UVCS) aboard the {\it SOHO} satellite
has observed very fast Coronal Mass Ejections (CMEs) associated with X-class
flares.  These events show spectral signatures different than those seen in 
most other CMEs in terms of very rapid disruption of the pre-CME streamer, very
high Doppler shifts and high temperature plasma visible in the [Fe XVIII]
emission line.   This paper describes three very similar events on 
21 April, 23 July and 24 August 2002 associated with X-class flares.
We determine the physical parameters of the pre-CME streamers and discuss
the geometrical and physical nature of the streamer blowouts.  In the
21 April event, the hot plasma seen as [Fe XVIII] is not related to the 
structure seen in [Fe XXI] by SUMER at lower heights.  It has the form
of a rapidly expanding fan, quite likely a current sheet.  In the August
event, on the other hand, the [Fe XVIII] is probably a bubble of hot
plasma formed by reconnection in the wake of the CME.  C III emission from
the July 23 flare is detected as stray light in the UVCS aperture.  It precedes
the hard X-ray brightening by about 2 minutes. 

\end{abstract}

\keywords{solar wind-- Sun:activity--Sun:corona--Sun:coronal mass ejections (CMEs)}

\section{Introduction}

\bigskip
Coronal Mass Ejections (CMEs) span a range in speed from below 100 to 2500 $\rm km~s^{-1}$ 
(e.g. St. Cyr et al. 2000; Gallagher et al. 2003).
They can occur with or without a powerful flare and with or without a prominence
eruption, but the fastest, most powerful CMEs seem to be associated with 
flares (e.g. Moon et al. 2002).  It is plausible that the largest flares and
the most energetic CMEs would arise from active regions where a great deal of
energy is stored in stressed magnetic structures.  It is therefore important
to investigate how CMEs associated with X-class flares differ from other CMEs,
especially as the fast CMEs can strongly affect the near-Earth environment.

A few CMEs were observed in the UV with Skylab and SMM (Schmahl \& Hildner 1977;
Fontenla \& Poland 1989).
Extensive observations of CMEs in the ultraviolet became possible with the launch
of the {\it SOHO} satellite, which carries 3 UV spectrometers.  These observations
show bright emission from relatively cool gas originating in ejected prominences
(e.g., Ciaravella et al. 1997, 2000).  More rarely, they show 
[Fe XVIII], [Fe XIX] and [Fe XXI] lines indicating temperatures 
above $6 \times 10^6$ K (Innes et al. 2002; Raymond 2002).  In a few 
cases, broad UV line profiles show coronal shock waves associated with Type II bursts
(Raymond et al. 2000; Mancuso
et al. 2002).  Doppler shifts exceeding 1000 $\rm km~s^{-1}$ have been observed,
and Doppler velocities sometimes reveal helical structures and rapid rotation (Ciaravella
et al. 2000; Pike \& Mason 2002).

Here we present UVCS (Kohl et al. 1995; 1997) observations of 3 very fast CMEs.  Their UV signatures were
quite different from those of other events we have observed.  All three show
transient brightenings in the [Fe XVIII] line and little or none of the cool
gas that usually dominates UVCS spectra of CMEs.  In all three cases the
pre-CME streamer is violently disrupted.  The O VI line profiles 
split into strongly red- and blue-shifted components, and the region of
split profiles along the spectrograph slit grows rapidly.  The components
of the split profiles are relatively narrow, indicating a lack of strong shock
heating.  

The three events were associated with X-class flares on 21 April, 23 July
and 24 August 2002. They are among the best-observed events in recent
history.  Thanks to the Max Millenium campaign,
EUV images and spectra from the {\it SOHO} and {\it TRACE} satellites
along with numerous ground-based radio and optical observations complement
the hard X-ray images and spectra from the {\it RHESSI} satellite.  The
21 April event and its effects on the upper atmosphere were also observed by
the {\it TIMED} satellite.   Strong Solar Energetic Particle events were
associated with the West limb events on 21 April and 24 August.  The July
event on the East limb was presumably not magnetically connected to the Earth.

The following sections describe the observations, then discuss the physical parameters
of the pre-CME streamers, the implications of the
split O VI profiles, the structure and evolution of the events in 3 dimensions
inferred from the Doppler shifts, and the nature of the [Fe XVIII]-emitting plasma.

\section{Observations}

All three events occurred in active regions targeted by
the Max Millenium campaign.  In all three cases, UVCS used the observing sequence designated
for Major Flare Watches.  The UVCS slit was placed 1.64 \RSUN\/ from Sun center
at the position angle directly above the target active region.  The 
100 micron slit width delivered a spectral resolution of 0.4 \AA\/ and a 
spatial resolution of 28\arcsec\/ in the radial direction.  In order to maximize spectral
coverage within telemetry limitations, the data were binned to 42\arcsec\/ along 
the slit length.  Five wavelength panels on the OVI detector 
covered the intervals 942-963, 966-980, 990-994, 997-1008 and 1023-1042 \AA .  
These intervals are chosen to include the lines Ly$\beta$, Ly$\gamma$ and Ly$\delta$
along with O I $\lambda \lambda$989, 991, [Si VIII] $\lambda \lambda$945, 949, 
[Si IX] $\lambda$950, C III $\lambda$977, N III$\lambda \lambda$990, 992, 
O VI $\lambda \lambda$1032, 1037 and [Fe XVIII] $\lambda$974.  The
Si XII $\lambda \lambda$499, 521 doublet is detected in second order, and the Fe XV
$\lambda$461 line is detectable if it is sufficiently bright.  

Data calibration is described by Gardner et al. (2002).  Because all these observations were taken with
a very narrow occulter width exposing only about 2.4 mm at the edge of the mirror,
we have corrected for degradation of that small part of the mirror (Gardner et al.
2002).  For the particular mirror and internal occulter positions chosen for
these observations there is a small contribution of stray light from the disk in the Ly$\beta$
line.  We account for this by measuring the C III $\lambda$977 intensity, which is pure stray
light, and multiplying by the ratio of Ly$\beta$ to C III measured on the disk (0.74;
Vernazza \& Reeves 1978).  This stray light contribution is under 10\% in the streamers but
can approach 30\% at the ends of the slit where the intrinsic Ly$\beta$ emission is weak.
For each event, a series of 120 second exposures was obtained with about 10 seconds
for readout between exposures.

We now briefly describe the three individual events.  Figure 1 shows a pair of LASCO C2
images for each eruption.  The contrast of the April and August events was enhanced by
wavelet analysis.  The images for the July event were part of a polarization sequence,
and they have not been enhanced.

\bigskip
{\it 21 April}: As part of the Max Millenium campaign, the UVCS instrument 
was pointed above Active Region NOAA AR 9906 at a position
angle PA=262$^\circ$ east of solar north.  The series of 120 second exposures began many hours before the
event and lasted until 01:55 UT.  At that time UVCS began its minisynoptic
program, which observes at 8 position angles around the Sun.  At the first of these
position angles, 270$^\circ$, a set of four 120 second exposures was obtained at 1.7 \RSUN\/
followed by 4 exposures at 1.52 \RSUN .  These exposures covered the wavelength
ranges 1023-1043, 998-1001 and 971-985 \AA .  The last of these also covered the
range 1208-1221 \AA\/ by way of the redundant mirror, so that it included the
Ly$\alpha$ line.
 
The GOES X1.5 class X-ray flare began at 00:59 UT and peaked at 01:31 UT while the
active region was located at latitude S14 and longitude W84.  It was accompanied
by Type II and Type IV radio bursts, a partial halo CME and a solar proton event.
The CME speed reached 2500 $\rm km~s^{-1}$, but at the
time it crossed the UVCS slit the speed was 1000 $\rm km~s^{-1}$ and the acceleration
near its peak value of 1500 $\rm m~s^{-2}$ based on a combination of TRACE, UVCS
and LASCO data (Gallagher et al. 2003).
RHESSI observations and TRACE images showed a rapidly growing arcade of loops beneath
a spiky structure (Gallagher et al. 2002).  The TRACE observing band includes both
strong Fe XII emission lines, probably responsible for the bright loops, and
an Fe XXIV line formed at about $2 \times 10^7$ K, that probably accounts for
the higher spiky structure.  The hard X-ray source moved upwards at a few $\rm km~s^{-1}$,
which fits in with the standard picture of double ribbon flares (see, for instance,
the review by Low 2001).
RHESSI data also showed gamma ray emission from the Earth's atmosphere due
to the impact of energetic particles (Share et al. 2002).
Simultaneous SUMER observations detected Doppler-shifted
emission in both very cool and very hot lines, C II to [Fe XXI] (Wang et al. 2002).
The C II emission showed a jet at the flare onset, while
the [Fe XXI] emission showed modest Doppler shifts from
a relatively slowly rising loop.  A shock from this event was recorded by the CELIAS/PM
on 23 April at 04:08 UT.

{\it 23 July}:  The target region AR 0039 was near the east limb
at latitude S13 and longitude E72 when an X4.8
flare occurred, beginning at 00:23 UT and reaching a peak at 00:29 UT.
It was accompanied by Type II and Type IV radio bursts.  LASCO reported a
fast halo CME whose leading edge was relatively faint and which showed
little bright emission.  The LASCO images indicate a very fast CME, though
the leading edge is poorly defined.  The leading edge may already be
beyond the edge of the C2 field in the first image in which the CME
is visible, giving a lower limit of about 1000 $\rm km~s^{-1}$.  Comparison
of the LASCO C3 images with the start time of the flare gives an
average speed of 2200 $\rm km~s^{-1}$.  The CME opening angle was nearly 180$^\circ$,
and there was no bright central core.
UVCS was pointed at PA=96.1$^\circ$ during the event. 
X-ray spectra from RHESSI show a departure from a pure power law, indicating
a combination of an increased column of ionized gas consistent with chromospheric
evaporation, though the photospheric albedo could also account for some of the
difference from a power law (Kontar et al. 2003).  RHESSI observations of the
511 keV line profile indicate positron annihilation at very high densities 
at transition region temperatures (Share et al. 2003).

{\it 24 August}: The target was AR 0069, the subject of a Major Flare Watch 
at a position angle of 260$^\circ$.  The X3.1 flare at S02, W81 
began at 00:49 UT and peaked at 01:12 UT, and it was accompanied by Type II and
Type IV radio bursts.  LASCO images show a fast, bright loop with some
trailing material.  LASCO C2 and C3 images indicate a speed of 1700 $\rm km~s^{-1}$
for the faint outermost loop.  The brighter emission that fills much of the
CME volume is moving about 2/3 as fast.  CELIAS/PM recorded a shock associated 
with CME on 26 August at 10:20 UT.
The UVCS observations began long before the X-ray flare and continued through 06:49 UT.

\subsection{Overview}

Figure 2 shows the intensity along the UVCS slit as a function of time for
each of the 3 events.  In each case a 65 minute interval is shown in the 
intensities of the O VI $\lambda$1032, Si XII $\lambda$499 and [Fe XVIII] $\lambda$974 lines.
The horizontal axis is the position along the UVCS slit.  (Note that North and South are 
reversed for the July 23 event, but slit center is still indicated by 0.) In all three 
cases relatively bright O VI emission originating in the active region streamer is seen 
near the center of the slit before the event.  Si XII is reasonably bright in the
pre-CME streamers on 21 April and 23 August.  [Fe XVIII] is present, but barely detectable,
in the 21 April and 24 August streamers.  

In the July 23 (and to a lesser extent the Aug. 24) event, a sudden brightening can be
seen across the entire length of the slit.  This is due to
increased scattering of photons from the solar disk due to the brightening
of the flare.  In the [Fe XVIII] and Si XII panels this scattered contribution 
is grating-scattered Ly$\alpha$.  As mentioned above, some stray light from the
solar disk is detected in C III $\lambda$977.  Before the flare it is detected
at a level of about $1.8 \times 10^8~\rm photons~cm^{-2}~s^{-1}~sr^{-1}$ averaged
over the length of the UVCS slit.  Between 00:22 and 00:26 UT the brightness
doubles, and it peaks at $8.6 \times 10^8~\rm photons~cm^{-2}~s^{-1}~sr^{-1}$
in the exposure lasting from 00:26;12 to 00:28:12 UT (Figure 3).  It then fades
over the course of about 15 minutes.  The C III brightening implies a large emission
measure of $10^5$ K gas during the time interval when RHESSI observed the
positron annihilation line originating in dense gas at transition region
temperatures (Share et al. 2003).  While the C III temporal profile supports the
idea of dense gas in the $10^5$ K range, it is interesting that the C III
brightening begins 2-6 minutes before the hard X-ray brightening reported
by Share et al., and the C III emission peaks in the exposure that spanned the 2 minutes 
before the hard X-ray rise.  The times given by the UVCS data files are updated monthly
to the SOHO spacecraft computer time, and they are accurate to about 15 seconds. 
The fact that C III brightening precedes the hard X-ray brightening
suggests that the dense transition region gas is heated by lower energy electrons
or by protons.

In the O VI lines the brightening is not stray light.
The excess O VI brightness above the pre-CME values shows a 4:1 
intensity ratio between the 1032 and 1037 \AA\/ lines, indicating resonance scattering
of O VI photons from the disk by O VI ions in the corona.  In the August event 
(and in the northern section of the July event), brightening in Si XII along the 
leading edge just after the scattering enhancement indicates enhanced local collisional 
excitation, and this would probably apply to the O VI emission as well.  We discuss below 
whether this results from shock compression or from denser material brought into the UVCS 
field of view from lower heights.

The scattered light components last for only a few exposures before the 
streamers are disrupted by the expanding CMEs, leaving only a few wisps of 
material in O VI and Si XII.  The edge of the
CME bubble along the slit can be seen as the concave upward emission rim in 
each event.  Small, transient [Fe XVIII]
features appear in the 21 April and 23 July events, in both cases lasting only
a few frames.  In the August 24 event, more diffuse transient emission appears
to occupy a significant fraction of the CME volume just behind the front.

Figure 4 displays the intensities of several lines against time to give a
quantitative version of some of the information in Figure 2.  In each case we
have averaged several spatial bins along the slit chosen to include the
brightest [Fe XVIII] emission.  The spatial regions chosen are  105\arcsec\/ south
to 105\arcsec\/ north, 147\arcsec\/ to 987\arcsec\/ north, 
and 315\arcsec\/ to 735\arcsec\/ south of slit center for the April, July and August
events, respectively.  The larger spatial averages were needed for the July and August
events because of their fainter emission.  

It is clear that the three events have similar time histories, though the evolution of
the faint July event is less clear.  In each case the brightness is
constant before the event.  The O VI emission suddenly brightens by 10\% to 50\%
for a few hundred seconds, then drops to 0.1 to 0.75 times the pre-CME value.  Si XII
brightens just after the O VI emission in the August event, but does not
brighten in the other two.  Had we chosen somewhat different spatial regions along the
slit, Si XII brightening would have appeared in all three.  In all three events, the 
Si XII dims rapidly in a manner
similar to the O VI emission.  In the July event, the Si XII and O VI brightnesses recover
about half an hour after the initial dimming.  In all three cases, a transient
brightening is [Fe XVIII] occurs just after the dimming in O VI and Si XII.

The steep drop in O VI and Si XII emission corresponds to disruption of the streamer.
The streamer gas is displaced by CME material of either lower density or of different ionization
state.  The CME plasma cannot be lower in ionization state, as that would show up as
bright emission in the Lyman lines, O VI or C III.  In the April and July events, higher
temperature gas is revealed by the [Fe XVIII] brightening, but only in a restricted
region along the slit. Only in the case of the August event does [Fe XVIII] emission fill
much of the region behind the leading edge.  We discuss the densities in more detail
below, but conclude that the sharp drop in intensities in the April and July events
results from a density drop, presumably corresponding to the CME void.  In the July
event a feature appears in Si XII and O VI at about 00:56 UT.  It is blue-shifted
by about 300 $\rm km~s^{-1}$.  Still later, at 02:17 UT, another small transient
feature appears, this time in the low temperature lines O VI and C III and blue-shifted
by about 150 $\rm km~s^{-1}$.  Both features
seem to be among the many approximately radial features seen in LASCO C2 images.

Finally, all three events show [Fe XVIII] emission, and this is fairly unusual among
CMEs observed by UVCS.  Bright, narrow [Fe XVIII] emission features in the wakes of 2
CMEs have been observed to last for many hours (Ciaravella et al. 2002; Ko et al. 2003),
and these have been attributed to the current sheets beneath expanding flux ropes predicted
by several CME models (e.g. Lin \& Forbes 2000; Amari et al. 2003; Webb et al. 2003;
Roussev et al. 2003).  A more diffuse
cloud of [Fe XVIII] emission appeared during an event on 26 October, 2000 (Raymond 2002).
In general, however, CMEs observed at heights above 1.5 \RSUN\/ show low ionization
material such as C III and O VI rather than high temperature gas.  In all three events,
the [Fe XVIII] appears after the disruption of the streamer seen in O VI and Si XII,
indicating that it is ``inside" the CME.   However, the morphologies of the [Fe XVIII]
emitting gas are different in the three events.  We will discuss the parameters of this
hot gas, its location within the CME structure and its physical origin in section 3.4.

\section{Analysis}

\bigskip
\subsection{Atomic Rates and Disk Emission}

Analysis of the observed line intensities requires a knowledge of the
excitation rates by collisions and by radiation from the solar disk.
For the excitation rates we use those of version 4.01 of CHIANTI
(Young et al. 2003), except that we use the rates of Raymond et al.
(1997) for the Lyman lines.  We use the ionization balance of
Mazzotta et al. (1998).  Disk intensities for the O VI doublet
and Lyman lines from Raymond et al. (1997) are multiplied by a
factor of 1.85 to account for the increase with solar activity
as estimated from the solar cycle dependences reported by 
Sch\"{u}le et al. (2000), McMullin et al. (2002) and Woods et al. (2000)
A somewhat larger disk flux might be
appropriate for these pointings directly above active regions,
which could increase the radiative excitation rates and the
derived densities by up to 20\% (Ko et al. 2002)

\bigskip
\subsection{Streamer Parameters}

In general the analysis follows the methods used by Raymond et al.
(1997), Ciaravella et al. (2002), Ko et al. (2002) and Uzzo et al.
(2003), but the CME watch campaign sacrificed spectral coverage 
for time coverage (using a single grating position) and spatial 
resolution.  As a result, the uncertainties in derived
temperatures and abundances are larger.

Figure 5 shows the intensity distributions of several lines along the
UVCS slit shortly before each of the CME events.  In each case we have
averaged 50 exposures to improve the signal for the [Fe XVIII] line.

The April and August streamers show single-peaked broad enhancements
in Ly$\beta$, while in the July streamer, Ly$\beta$ is relatively constant
along the UVCS slit.  The O VI lines, on the other hand, show double-peaked
structures in all three cases.  (The O VI 1037 line is not plotted because
it is close to 1/3 the intensity of 1032 everywhere.) The structures are
reminiscent of quiescent equatorial streamers at solar minimum
(Raymond et al. 1997; Uzzo et al. 2003).  In those cases
the temperatures and ionization states were fairly constant across
the streamers, so that the variation in O VI to Ly$\beta$ ratio
was attributed to changes in the oxygen abundance, with O depleted
within the central streamer core.  In the three streamers considered
here, variations in the Si XII to [Si VIII] ratio indicate that the
ionization state varies across the streamer.  Part of the central
dip in O VI in the April streamer and the weakness of the O VI
peak at -1 arcmin in the July streamer clearly results from higher
average ionization states.

We have extracted spectra of 2 or 3 regions along the slit for each
streamer for further analysis.  For April 21 we choose the bright
O VI peak near -7\arcmin\/ and the O VI minimum near -1\arcmin.  For
the July streamer we take the O VI and Si XII peak at -10\arcmin\/
and the O VI dip at -4\arcmin, and for the August streamer the
O VI peak at -8\arcmin, the Si XII peak at -1\arcmin\/ and the
O VI dip at -4 arcmin.  The spectra are presented in Table 1.
Extraction intervals and derived parameters are shown in Table 2.  

Table 2 also shows the Ly$\beta$ and Ly$\gamma$ intensities
corrected for stray light (see section 2).  We also show
the intensity ratio of the O VI doublet, which gives the ratio 
of the collisional contribution (2:1) to the resonance scattering 
component (4:1).  Provided that the outflow speed is small, which 
is a good approximation at this height (Strachan et al. 2002),
the ratio of collisional to radiative contributions gives a density
(Kohl \& Withbroe 1982), which is also
listed in Table 2.  Both the O VI doublet ratios and the densities
are quite similar among all the features observed.  The densities
are intermediate between those of solar minimum equatorial streamers
(Strachan et al. 2002) and densities at a similar height
above a very hot active region (Ko et al.  2002).  They are
similar to densities at 1.6 \RSUN\/ in streamers at solar
maximum (Parenti et al. 2000).

Crude estimates of average temperatures were
obtained from the ratios of [Si VIII] to Si XII.  Not surprisingly,
the derived temperatures fall between log T = 5.9, the emission peak of
[Si VIII] and log T = 6.3,  the peak of Si XII emissivity.
While the values are similar to streamer temperatures near $1.3-1.6\times 10^6$ K
derived with much more extensive sets of emission lines (e.g.
Raymond et al. 1997; Feldman et al. 1998; Parenti et al. 2000),
the trend among the features observed is more significant than
the numerical value for any feature.  The width of Ly$\beta$
puts an upper limit of about $1.8 \times 10^6$ K on the
ion temperature.  However, faint [Fe XVIII] emission is
detectable in the April 21 streamer leg and at all 3 positions
in the August 24 streamer.  This indicates the presence
of a small amount of gas at log T $\ge$ 6.4.  The limited spectral range
of this observing sequence makes determination of a differential
emission measure curve possible, but ambiguous.  In general,
flat differential emission measures over the range log T = 6.1
to log T = 6.3 can match the spectra without [Fe XVIII], and a somewhat
smaller amount of gas at log T $\ge$ 6.4 provides the [Fe XVIII] emission.
Table 2 also shows the emission measures derived from the Ly$\beta$
fluxes assuming the densities in Table 2 and an average temperature
Log T = 6.2.  In each case, combining the density with the emission
measure yields a thickness of the emitting regions of order 0.5 \RSUN .

We have also used the ratios of O VI to Ly$\beta$ to estimate the 
absolute abundance of oxygen.  Because we do not have Ly$\alpha$
intensities, we cannot separate the collisional
and radiative components of Ly$\beta$, as was done by Ciaravella
et al. (2002) and Ko et al. (2002).  The Ly$\beta$ to Ly$\gamma$
ratio could be used in principle, but this ratio is much less sensitive
to the radiative to collisional ratio, and there are larger uncertainties
in both measurements and atomic rates.  Therefore, we use the density
and temperature estimates from Table 2 to compute total emissivities
for the Ly$\beta$ and O VI lines.  That means that  abundance results 
shown in Table 2 are not as accurate as those from studies designed to 
measure abundances.  Nevertheless, they show a distinct difference from both solar
minimum streamers and hot active region streamers studied at this
height (Raymond et al. 1997; Ko et al. 2002; Ciaravella et al.
2002; Uzzo et al. 2003) where the oxygen abundance is
roughly 1/3 photospheric in the streamer legs and 1/10 photospheric
in the streamer cores.  Here the abundances are about 3 times
larger, and similar to those derived by Parenti et al. (2000) for
a streamer in June 2000.  The low abundances in the other streamers were attributed
to gravitational settling, and it was suggested that coronal
and chromospheric material must be mixed on about a 1 day time scale
to account for the observed depletions.  The higher abundances
in the streamers associated with powerful flares and CMEs may
indicate mixing with chromospheric material on a shorter time scale
due to the high activity levels in these regions.  Abundances
for Si and Fe are more difficult to determine because the lines
of those elements are formed above and below the average temperature,
so they depend upon the details of the emission measure distribution.
In general, however, they suggest First Ionization Potential
(FIP) enhancements of about 3, a typical streamer value (e.g.
Raymond et al. 1997; Parenti et al. 2000; Ko et al. 2002; Uzzo et al. 2003)

\bigskip
\subsection{Streamer Blowout}

\bigskip
Figure 6 shows the spectral region near the O VI $\lambda$1032 emission lines
for the April and August CMEs at six
times; just before the eruption and at 5 times over the following 15 minutes.
The July event is not shown because, although it appears to be similar,
the O VI intensities are smaller by roughly the ratio of the pre-CME
streamer intensities, and therefore the signal-to-noise is poor.
The Si XII $\lambda$499 line in the August event evolves in a manner
nearly identical to that of the O VI lines, but because the count rate
is smaller than that of O VI by a factor of 4, the signal-to-noise is 
not as high.  Ly$\beta$ also seems to show the same disruption, but the 
count rate is even smaller.

In both CMEs, the region near the slit center is the
first to be affected, as emission at line center dims 
and strongly red-shifted and blue-shifted components
appear.  The disturbance spreads rapidly along the slit, and Doppler shifts
reach +500 and -850 $\rm km~s^{-1}$ in the 21 April event, or -500 and +810
$\rm km~s^{-1}$ on 24 August.  On 21 April on the red side, the Doppler shift
peaks about 7 minutes after the red-shifted component appears, then declines
to roughly +150 $\rm km~s^{-1}$.  On the blue side, the Doppler shifts seem
to reach asymptotic speeds of 750 to 850 $\rm km~s^{-1}$, and the emission
simply fades away.  The fact that blue-shifts are larger than red-shifts on 21 April
implies a skewness towards the Earth, perhaps related to strength of the
observed near-Earth effects.

The long slit spectra in Figure 6, particularly the spectrally narrow red-
and blue-shifted peaks, imply that the O VI emitting gas is concentrated
in relatively thin sheets.  These are presumably the higher oxygen abundance
regions seen as
the brightness peaks in the intensity distributions along the slit.  The
dimming at line center corresponds to displacement of the streamer by
either the oxygen-depleted closed field core or the low density CME void.
Doppler dimming of O VI in the moving plasma further reduces the
O VI brightness (Noci, Kohl \& Withbroe 1987).

For about the first 500 seconds, the disturbance spreads along the slit 
at a rate of about 500 $\rm km~s^{-1}$ in each direction, comparable to the 
speeds of the CME fronts at this height.  The 1000 $\rm km~s^{-1}$ total motion
along the slit is a little smaller than full range of the Doppler shift velocities.
The Doppler shifts are made up of two contributions; the expanding motion of the
magnetic structure along the line-of-sight (LOS) and the LOS component of plasma
motion along the field.  The similarity of the Doppler expansion and the expansion
along the slit suggests that, at least at this stage, outflow along the
field lines is a smaller contribution than transverse expansion of the magnetic
structure.  The LASCO C2 images of
the July and August events give a strong impression of transverse expansion,
while the April event has more of the appearance of a cloud of ejected plasma.

A slab of the corona that is 0.25 \RSUN\/ thick having the density derived above 
and lateral dimensions of about 1 \RSUN\/ would weigh 
about $10^{15}$ grams, or the mass of a modest CME (e.g. Hundhausen 1997).  Thus the disrupted streamer
could account for much of the CME mass. However, as pointed out above, it is
not clear how much of the displaced mass is ejected and how much is merely
pushed aside.

The generally narrow O VI profiles also indicate that O VI is not heated to high 
temperatures.  Single-peaked line profiles about 900 $\rm km~s^{-1}$ wide have 
been observed in CME-driven shocks (Raymond et al. 2000; Mancuso et al. 2002).
It is difficult to assign definite line widths to the red- and blue-shifted 
components of O VI in the X-class flare related  events, but they are no more
than 100-200 $\rm km~s^{-1}$ wide, and some of that width must be attributed to bulk
motion of the plasma.  The absence of shock heating is in keeping with the appearance of gradual
acceleration as the Doppler shifts increase to their final values over
the course of several exposures.  The apparent lack of a shock
suggests a fast mode speed faster than the CME speed.  With the densities
derived above, a fast mode speed above 1000 $\rm km~s^{-1}$ requires 
$B \ge 1$ Gauss.  This is comparable to the field strength at this height
in solar minimum streamers (e.g. Li et al. 1998), and active region streamers
should have stronger fields.

\subsection{Hot Plasma}

\bigskip
The three CMEs associated with X-class flares are unusual among CMEs
observed by UVCS in that they show [Fe XVIII] emission, which peaks
at log T = 6.7-6.8 in ionization equilibrium. UVCS
spectra of some events show bright, narrow [Fe XVIII] emission
features that last many hours \cite{ciaravella02, kob}.  These
are identified with the current sheets predicted by many flux rope models
of CMEs (e.g. Lin \& Forbes 2000; Manchester 2003; Amari et al. 2003;
Roussev et al. 2003).
It is perhaps surprising that [Fe XVIII] emission is not observed
more frequently, in that an abnormally high ionization state is 
a common signature of CME material measured {\it in situ} \cite{lepri}.
Thus these events may help to link coronal and interplanetary 
observations of CMEs.

The spatial distributions of the [Fe XVIII] brightness are different
in the 3 events.  The April event shows a relatively bright, sharply
defined feature about 2\arcmin\/ across.  The bright emission is
visible in 4 exposures (8 minutes) and more diffuse emission 
persists for another two exposures.  The [Fe XVIII] emission is close
to, but not cospatial with, a narrow feature seen in O VI and Si XII that
persists for a longer time (Figure 2).  The [Fe XVIII] appears
three exposures (6 minutes) after the O VI profile begins to split,
placing it well inside the CME front.  The first appearance of [Fe XVIII]
is about 15 minutes after the flare onset, well before the flare peak.

Only the April event has high enough signal-to-noise to yield a good
line profile.  This is shown in Figure 7.  The line width is about
1.5 \AA\/ or 450 $\rm km~s^{-1}$.  It could be comprised of
separate peaks near +150 and -210 $\rm km~s^{-1}$, but examination
of the individual exposures indicates that it is a single broad
profile.
 
During the July eruption, only
a brief episode of [Fe XVIII] emission occurs, lasting for two exposures
near the northern end of the slit.  However, a narrow emission region
does appear later at about 01:55 UT, and it persists for about 1/2 hour.  This
emission may be related to the current sheets described by 
Ciaravella et al (2002) and Ko et al. (2002b).  It appears sooner after
the eruption and lasts for a shorter time than the other UVCS current sheets
or the white light current sheets described by Webb et al. (2003).  This
is probably related to the very high CME speed.  The lengths of current
sheets are expected to depend on CME speed and on background coronal
density (Lin 2002), with a high CME speed implying a short duration for the
current sheet.

The August event shows
a much more diffuse [Fe XVIII] region extending perhaps 15\arcmin\/
along the slit.  As in the April event, the [Fe XVIII] peaks several
exposures after the O VI and Si XII peaks, appearing as a shell or
loop inside the expanding O VI volume.  In the August event,
the O VI and Si XII intensities begin to fade at 01:06 UT when the
CME removes streamer material from the line of sight.  The transient
feature appears at 01:16 UT, reaching
$0.4, 7.0~\rm and~2.7\times 10^{9}~\rm photons~cm^{-2}~s^{-1}~sr^{-1}$
in  Ly$\beta$, Si XII and [Fe XVIII], respectively.
For solar photospheric abundances and ionization equilibrium, the line ratios
imply 6.7 $\le$ log T $\le$ 6.9 and an emission measure of 
$2 \times 10^{25}~\rm cm^{-5}$.  If the structure is more or less
cylindrical and the transverse scale is about 0.5 \RSUN , the 
emission measure and length scale imply a
density near $2 \times 10^7~\rm cm^{-3}$.  If the emitting region is  
an arcade seen end-on, the density could be somewhat smaller.

There are two plausible interpretations for the observed hot gas.
Since Fe XVIII is sometimes seen in X-ray spectra of active regions
(e.g., Rugge \& McKenzie 1985), the [Fe XVIII] might be produced 
by ejected active region material in the form of 
expanding loops.  On the other hand, there is obviously strong 
impulsive heating in these events, and the [Fe XVIII] may arise
from cooler gas that is heated by reconnection.  In particular,
current sheets are predicted either above the main CME volume, for instance in
the breakout model of Antiochos et al. (1999), or beneath the 
main CME magnetic flux rope in both breakout model and flux 
rope models such as those of Lin \& Forbes (2000) and Amari 
et al. (2003).  Hot plasma may be found in the current sheet
itself or on the magnetic field lines that have undergone
reconnection.  The ``breakout" model of Antiochos et al. (1999)
predicts that such plasma forms the flux ropes detected in 
interplanetary space as reconnection transforms a sheared
arcade into a helical structure.  The model of Lin \& Forbes (2000)
starts with a pre-existing flux rope, but reconnection
forms a ``bubble" of heated plasma of comparable size around it
(Lin, Raymond \& van Ballegooijen 2003).  

In all three cases the [Fe XVIII] emission appears after the 
onset of Doppler splitting of the O VI lines, implying that
it lies inside or beneath the main CME volume and cannot
arise in an overlying current sheet that initiates the breakout.
This does not imply that such a current sheet does not
exist, only that is not bright enough to be detectable.

In the April event, the [Fe XVIII] emission is concentrated in
a feature as small as 84\arcsec\/ along the slit.  This is most
naturally interpreted as a loop seen end-on, as one leg of a
loop, or as a current sheet seen end-on.  The broad line profile
roughly centered on zero velocity argues against the interpretation
in terms of a single leg of a loop, so the structure must be seen
end-on.  If the profile were interpreted as two peaks at -150 and
+210 $\rm km~s^{-1}$ an expanding loop would be the natural interpretation.
However, examination of the individual spectra shows that each
spectrum has a broad profile with blue-shifted or red-shifted
emission dominating in different exposures. This
implies that a fan of emission seen nearly edge-on
is the natural interpretation, favoring the idea that the 
hot gas resides in a current sheet.  A 1000 $\rm km~s^{-1}$
upward motion and $\pm 20^\circ$ opening angle in the plane of
the fan would account for the line profile.  The duration of about 800
seconds implies a height of just over 1 \RSUN\/ if the gas is
moving at 1000 $\rm km~s^{-1}$. The [Fe XVIII] emitting
gas appears in the fourth of the exposures that show splitting
in the O VI lines, at about the time that the maximum redshift
was seen but before the blueshift reached its maximum.  This
places the top of the [Fe XVIII] curtain about 1 \RSUN\/ behind
the top of the CME, perhaps near the bottom of the flux rope
described in the Lin \& Forbes (2000) model.  

An interesting possibility
within the context of current CME models is that the [Fe XVIII]
comes from the region where plasma accelerated in the current
sheet encounters the trailing edge of the flux rope.  The plasma
is expected to be moving at a few hundred $\rm km~s^{-1}$ relative
to the CME structure, and it is compressed and heated when it
encounters an obstacle.  The [Fe XVIII] surface brightness of 
$10^{9}~\rm photons~cm^{-2}~s^{-1}~sr^{-1}$ in a region as small
as 82\arcsec\/ can be combined with the emissivity of the line
and a depth estimate of 0.5 \RSUN\/ to give a density of about $10^7~\rm cm^{-3}$.
If the sheet is thinner than about 0.1 \RSUN\/ the density
may be higher.  The density results from compression as material
flows into the current sheet.  This interpretation is compatible
with either the breakout or flux rope models.

In the August event the curved morphology of the [Fe XVIII]
emission just inside the O VI and Si XII emission (Figure 2)
strongly suggests expanding hot plasma driving the cooler gas
in the upper regions of the streamer.  The density is a
factor of 5 higher than that of the pre-CME streamer,
a marginally higher density contrast than the factor of 4
for a strong shock.  The idea of a shock
producing the [Fe XVIII] while the O VI split profiles show the
absence of shock heating appears contradictory,  However,
streamer cores are generally believed to be high $\beta$
plasma (e.g. Li et al. 1998; Suess \& Nerney 2002; Endeve,
Leer \& Holzer 2003), while the streamer legs are lower
$\beta$.  Thus the disturbance could be super-Alfv\'{e}nic
in the core of the streamer, but sub-Alfv\'{e}nic in the
legs where O VI is formed.  Thus a shock interpretation is
possible in principle.  However, to the extent that
line profiles can be seen in the data the [Fe XVIII] feature
appears to be fairly narrow, indicating a lack of shock heating.

Therefore, the [Fe XVIII] in the August event is primarily denser
gas rising from below rather than compression of existing
coronal gas.  The hot plasma could lie on an arcade of expanding
loops directly tied to the photosphere.  However, the first
LASCO image showing the CME gives a clear impression that the
dense gas (and it was shown above that the [Fe XVIII] gas is
dense) takes the form of a helical flux rope.  The flux rope
in models having a pre-existing flux rope has low density
except perhaps in a cold prominence at the bottom of the magnetic
coils, so that most of the flux rope is identified with the
CME void (e.g. Gibson \& Low 1998).  Thus in the context of the flux
rope models the [Fe XVIII] appears to be the ``bubble" of
hot material that forms around the flux rope as reconnection
cuts off field lines and heats and injects additional plasma from the
current sheet (Lin et al. 2003). The
existence of this bubble is required by the standard flux
rope models (e.g. Lin \& Forbes 2000; Amari et al. 1999, 2003,
Roussev et al. 2003), but detailed predictions for its density and
temperature are not yet available.  It is not clear whether the
``bubble" can be distinguished from the pre-existing flux rope
by either magnetic characteristics (e.g. pitch angle) or by
plasma composition.  Similarly, the ``breakout"
model predicts the formation of a flux rope by reconnection
near the base of a sheared arcade, but the detailed properties
of the plasma are not yet predicted by the model.  In either type
of model, the plasma properties may depend on the details of
magnetic configuration and plasma characteristics in the 
pre-CME structure.  

We note that no obvious current sheet is seen behind the main body of
the August CME.  This is probably due to a partly face-on viewing angle,
which reduced the surface brightness of the thin structure below
our detection limit.  Thus some of the difference between
the April and August events lies in the accident of viewing geometry.

We close the discussion of the hot plasma by noting that the ionization
and recombination times of Fe XVIII at the densities derived above are about
3000 seconds, so the gas may not be in ionization equilibrium.
The ionization state probably freezes in when the density
is about an order of magnitude higher, roughly half way between
its ejection point and the height of the UVCS slit.  While the
excitation rate of the [Fe XVIII] line is relatively insensitive to
temperature, comparison of the [Fe XVIII] emission with X-ray
observations could determine whether the gas is rapidly cooled by
adiabatic expansion or rapidly heated by reconnection.  It is unfortunate
that there are no simultaneous soft X-ray spectra at 1 keV
energies.

\section{Summary}

We have observed 3 fast CMEs associated with X-class solar flares. 
A summary of the properties of the flares and CMEs is provided in
Table 3.
All three events show splitting of the O VI profiles during streamer
disruption and transient [Fe XVIII] emission.  Both features are
unusual among CMEs studied so far by UVCS.  The CMEs also show 
no evidence for the cool prominence material that usually dominates
UVCS spectra of CMEs.  The existence of a substantial
amount of highly ionized gas corresponds well to measurements of the
ionization state of ICMEs measured {\it in situ} \cite{lepri}.
While 3 events constitute a small sample, it seems that X-class
flares are associated with very fast CMEs which contain little
cool filament material.

The O VI lines provide a clear look at streamer disruption in 3D,
with line-of-sight speed comparable to the speed at which the
disturbance moves along the slit, both speeds are about 1/2 the speed
of the CME leading edge in the plane of the sky.  The temperature,
density and ionization state of the pre-CME streamers are similar
to those observed above hot, bright active regions.  The streamer
masses are comparable to typical CME masses, so the pre-CME
conditions provide a useful comparison for {\it in situ} measurements
of these quantities in the CMEs and in the associated energetic
particles.  However, it appears that the shocks had not formed at
1.6 \RSUN\/ due to fairly strong magnetic fields.

In the 21 April event, a small transient feature seen in [Fe XVIII]
is most easily identified with the current sheet predicted by CME
models.  In the 24 August event the more diffuse [Fe XVIII] emission
is apparently hot plasma that drives the expansion of the streamer
legs seen in O VI.  The [Fe XVIII] plasma probably comprises the
``bubble" of reconnected field lines that surround the original
flux rope (Lin et al. 2003) or the reconnected field lines that
are forming the flux rope (Antiochos et al. 1999).

\bigskip
This work is based on observations with the UVCS and LASCO instruments
aboard the SOHO satellite.  We thank the instrument teams and the
SOHO flight operations team.
The work benefited from workshop organized by the TIMED team
at APL.  The analysis was supported by NASA Grants  NAG5-11420 and NAG5-12827 to the Smithsonian 
Astrophysical Observatory.  The authors thank L. Gardner, J. Kohl, R. Wu, T. Forbes and H. Hudson for
especially helpful discussions.

\clearpage
 
\onecolumn

\begin{table}
\begin{center}
\centerline{Table 1}
 
\vspace*{5mm}
\centerline{ Line Intensities ($10^8~\rm photons~cm^{-2}~s^{-1}~sr^{-1}$)}

\vspace{5mm}
\begin{tabular}{| l c r r r r r r r |}
\hline \hline
ION & $\lambda$ & \multicolumn{2}{c}{April 21 } & \multicolumn{2}{c}{July 23} & \multicolumn{3}{c|}{August 24} \\
    &            &$~~$  Leg  &  Core  &$~~$  Leg &  Core  &$~~$  Leg1  & Core &  Leg2  \\
\hline
 Si VIII  &  $~$944.4  &    9.1  &    7.2  &    1.6  &    0.9  &    3.3  &    0.6  &    0.0  \\
 Si VIII* &  $~$949.4  &    3.3  &    2.5  &    0.7  &    0.3  &    1.1  &    0.3  &    0.4  \\
 Si IX*   &  $~$950.2  &    6.8  &    5.5  &    1.7  &    1.5  &    2.7  &    1.1  &    0.7  \\
Ly$\gamma$&  $~$972.5  &    3.2  &    3.8  &    1.4  &    1.6  &    2.0  &    2.6  &    2.8  \\
 Fe XVIII &  $~$974.8  &    0.7  &    0.0  &    0.0  &    0.0  &    1.5  &    0.9  &    0.2  \\
C III     &  $~$977.0  &    1.1  &    1.3  &    2.8  &    3.2  &    1.7  &    2.2  &    2.3  \\
Si XII    &  $~$499.4  &$~~$278.0  &  322.0  &$~~$  116.4  &   93.8  &$~~$  224.2  &  300.0  &  296.6  \\
Ly$\beta$ &    1025.7  &   18.2  &   24.4  &    8.9  &    8.9  &   14.1  &   16.9  &   12.0  \\
 Fe X     &    1028.0  &   10.2  &    8.1  &    2.5  &    2.0  &    3.9  &    1.2  &    0.9  \\
O VI      &    1031.9  &  652.0  &  506.0  &  202.0  &  173.0  &  312.0  &  165.0  &  182.0  \\
O VI      &    1037.6  &  212.0  &  168.0  &   67.4  &   54.6  &  101.0  &   53.6  &   60.6  \\
Si XII    &  $~$520.7  &  121.9  &  128.1  &   53.6  &   45.3  &  107.9  &  134.6  &  126.6  \\
\hline
\end{tabular}
\end{center}
* Blended with Ly$\delta$
\end{table}

\clearpage

\begin{table}
\begin{center}
\centerline{Table 2}
 
\vspace*{5mm}
\centerline{ Derived Quantities and Streamer Parameters}

\vspace{5mm}
\begin{tabular}{| l c r r r r r r r |}
\hline \hline
    &           & \multicolumn{2}{c}{April 21 } & \multicolumn{2}{c}{July 23} & \multicolumn{3}{c|}{August 24} \\
    &            &$~~$  Leg  &  Core  &$~~$  Leg &  Core  &$~~$  Leg1  & Core &  Leg2  \\
\hline

Lower Edge$^1$ & &   -8.75$^\prime$ & -3.15$^\prime$ & -11.55$^\prime$ & -6.65$^\prime$ & -10.85$^\prime$ & -5.95$^\prime$ & -2.45$^\prime$ \\
Upper Edge$^1$ & &   -4.55$^\prime$ & +0.35$^\prime$ & -7.35$^\prime$ &  -2.45$^\prime$ &  -5.95$^\prime$ & -2.45$^\prime$ & +1.75$^\prime$ \\
Ly$\beta_{corr}^2$ & &     17.4        & 23.5         & 6.85         &  6.55         & 12.9          & 15.3         & 10.3 \\
Ly$\gamma_{corr}^2$ & &    3.02        & 3.60         & 0.88         & 1.06          & 1.75          & 2.24         & 2.36 \\
O VI ratio & &    3.07        & 3.01         & 3.00         & 3.17          & 3.09          & 3.08         & 3.00 \\
$I(1032)_{coll}^2$ & & 197.6  & 166.4 & 67.3 & 44.9 & 92.3 & 49.6 & 60.7 \\
$\rm n_e^3$ &  &    4.8         & 5.3          & 5.4          & 3.9           & 4.5           & 4.6          & 5.4  \\
$\rm T_6^4$ & & 1.34 & 1.37 & 1.40 & 1.42 & 1.41 & 1.60 & $>$ 1.65 \\
$[$O/H$]^5$   & & 1.06 & 0.60 & 0.81 & 0.78 & 0.70 & 0.31 & $>$0.48  \\
EM          & & 10.4   & 15.2 & 4.6  & 3.3  & 7.3  & 8.8  & 6.9    \\
\hline
\end{tabular}
\end{center}
$^1$ Measured from 0 in Figure 5

$^2$ $\rm 10^{8} ~Photons~cm^{-2}~s^{-1}~sr^{-1}$

$^3$ $10^6~\rm cm^{-3}$

$^4$ $10^6$ K; derived from Si XII/Si VIII ratio

$^5$  Oxygen abundance relative to photospheric

$^6$  Emission Measure $10^{24}\rm ~cm^{-5}$
\end{table}

\clearpage
 
\begin{table}
\begin{center}
\centerline{Table 3}
 
\vspace*{5mm}
\centerline{ Summary of CME properties}
 
\vspace{5mm}
\begin{tabular}{| l c c c |}
\hline \hline
  & April 21 & July 23 & August 24 \\
\hline
Flare Class  &  X1.5  & X4.8  & X3.1 \\
CME speed*     & 2500   & 2200  & 1700 \\
$\rm V_{Doppler}$*& -850 to +500 & ? & -500 to +810 \\
Energetic Particles  & Y  & N  & Y  \\
IP Shock             & Y  & N  & Y  \\
511 keV line         & ?  & Y  & ?  \\
Flare C III          & N  & Y  & Y  \\
Diffuse [Fe XVIII]   & N  & N  & Y  \\
Sharp [Fe XVIII]     & Y  & Y  & N  \\
Shocked O VI         & N  & ?  & N  \\

\hline
\end{tabular}
\end{center}
*  $\rm km~s^{-1}$
\end{table}

\clearpage

\begin{figure}
\plotone{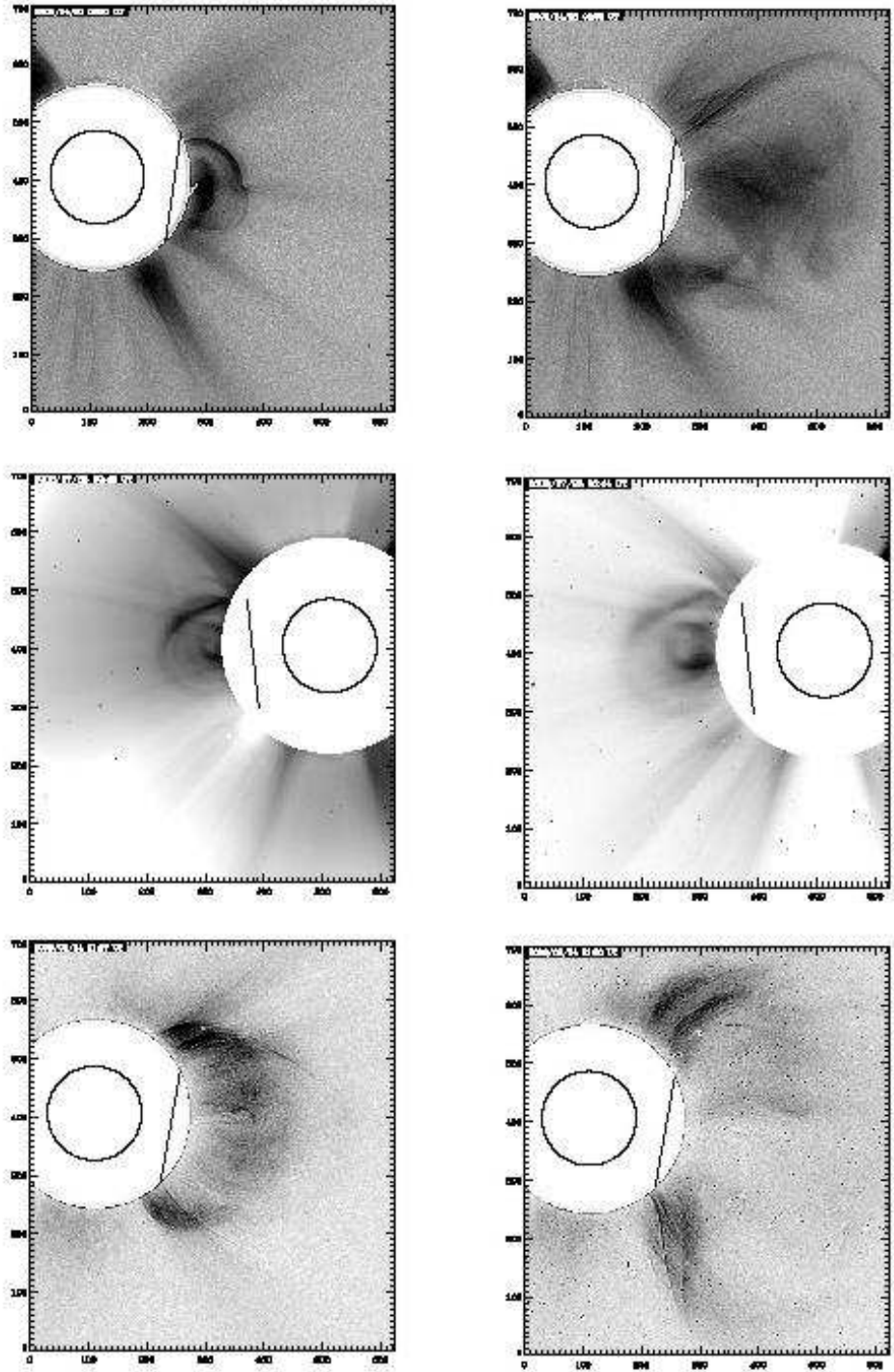}
\caption{Figure 1.  LASCO C2 images of the April (first row; 01:27 and 01:50 UT), 
July (second row; 00:42 and 00:44 UT) and August (third row; 01:27 and 01:50 UT)
CMEs with the UVCS slit positions superposed.
A wavelet analysis has been used to enhance the detail in the
April and August images.  The original images are shown for the July
event.  In all cases the tick marks show LASCO C2 pixels (11.4\arcsec ). \label{fig1}}
\end{figure}

\clearpage

\begin{figure}
\plotone{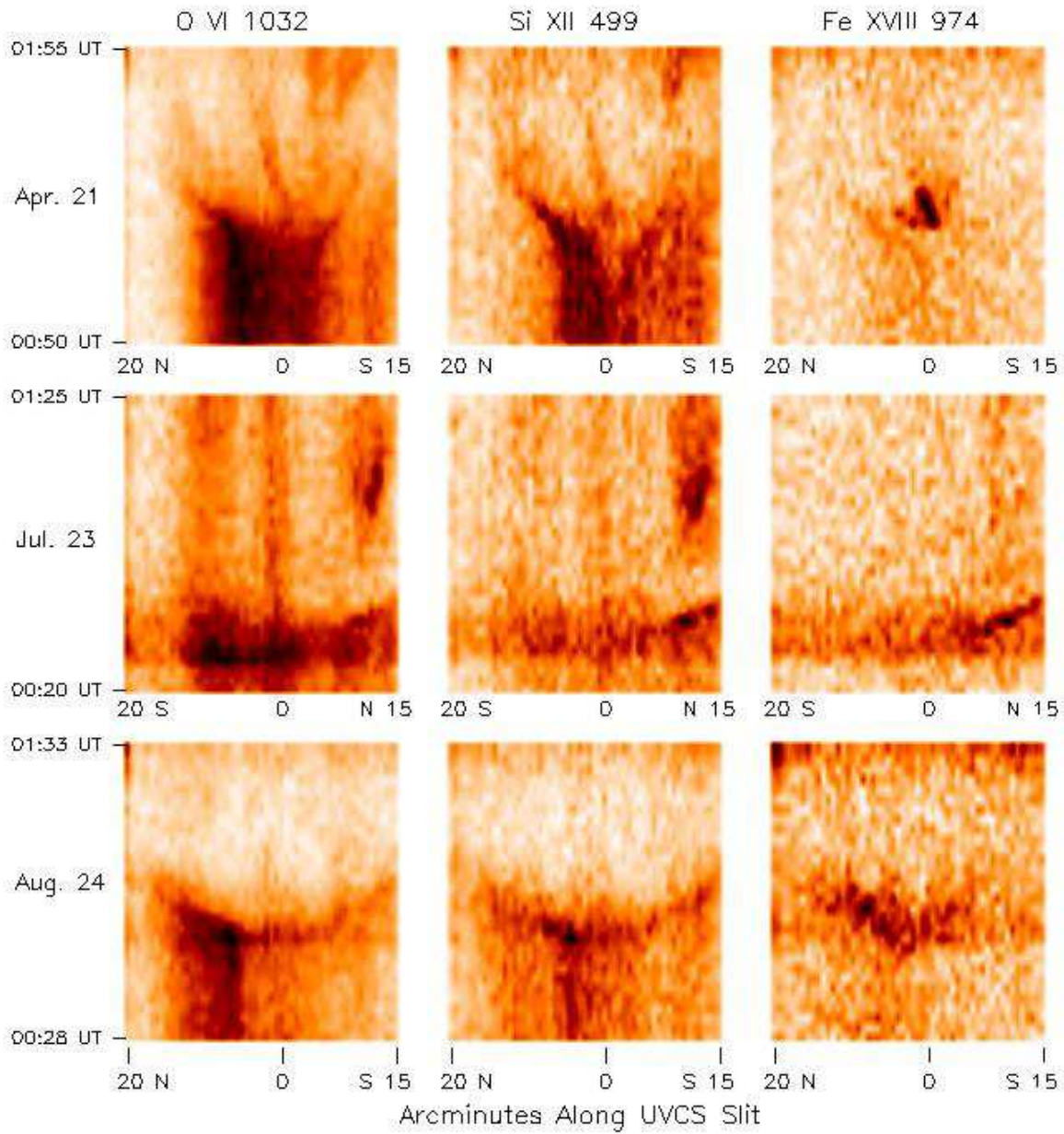}
\caption{Figure 2.  Intensities of the O VI $\lambda$1032, Si XII $\lambda$499
and [Fe XVIII] $\lambda$974 lines along the
UVCS slit (horizontal axis) and time (vertical axis).  The time interval
for each event is shown on the left, and time progresses
upwards.  The pre-CME streamer can be seen at the bottom of each panel.
Note that the North and South ends of the slit are reversed for the 23 July event.}
\end{figure}
 
\clearpage

\begin{figure} 
\plotone{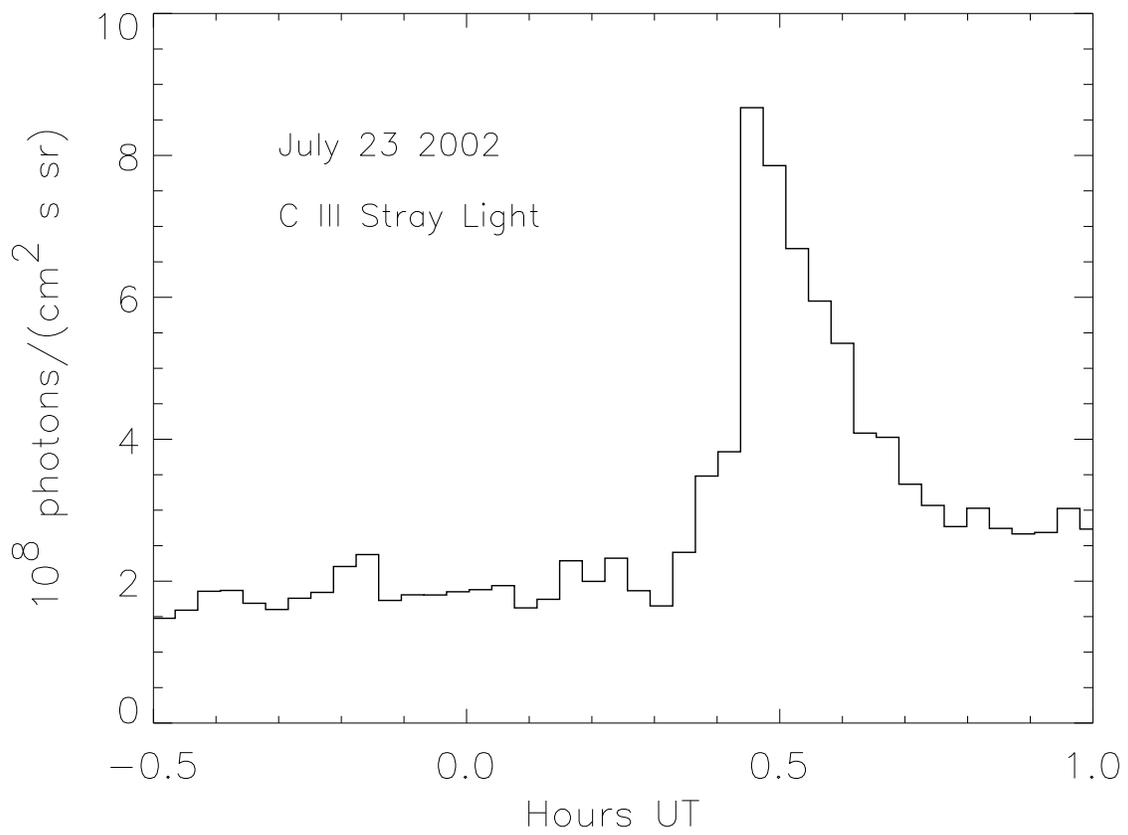}
\caption{Figure 3.  Time series showing the average stray light intensity of the
C III $\lambda$977 line.}
\end{figure}

\clearpage

\begin{figure}
\plottwo{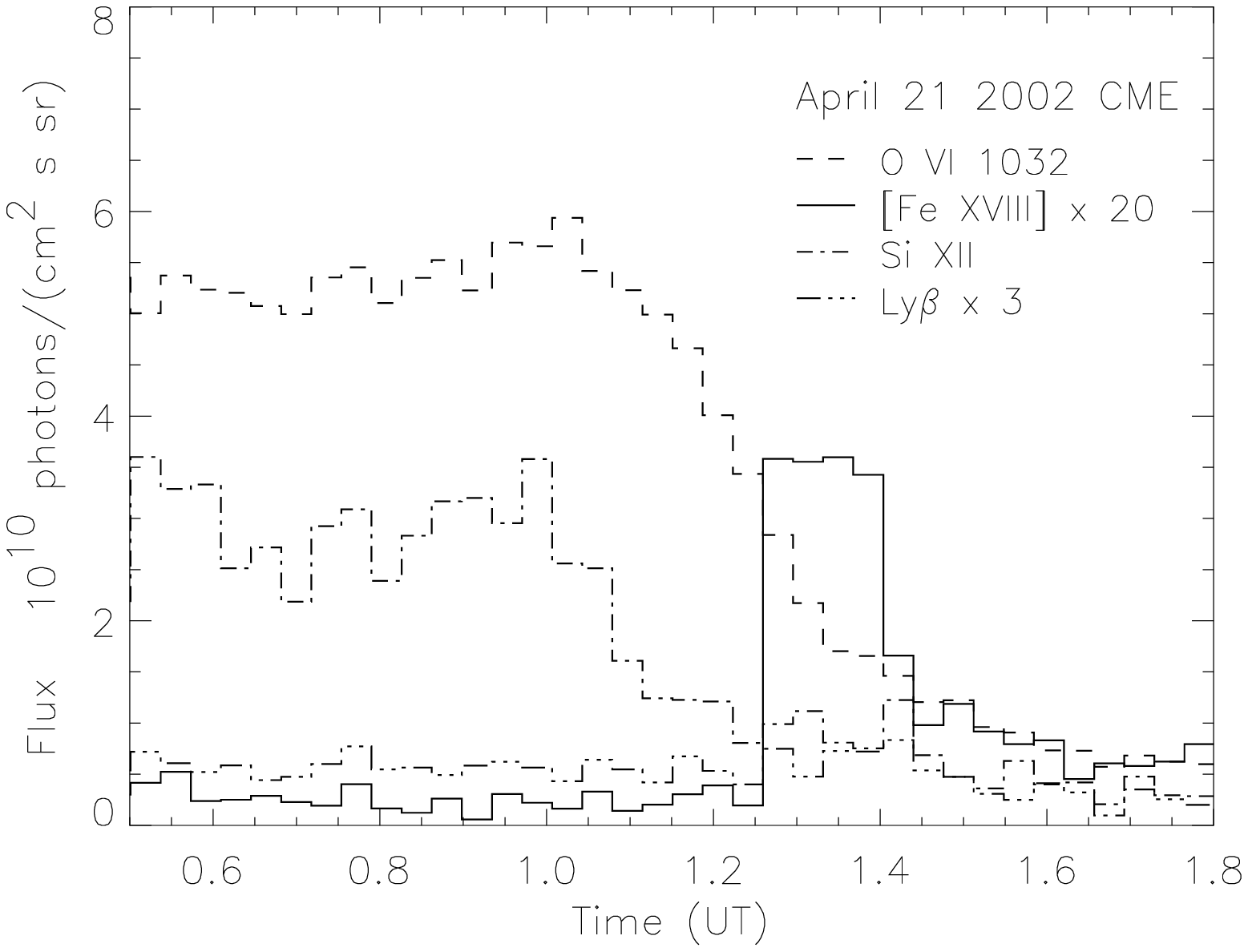}{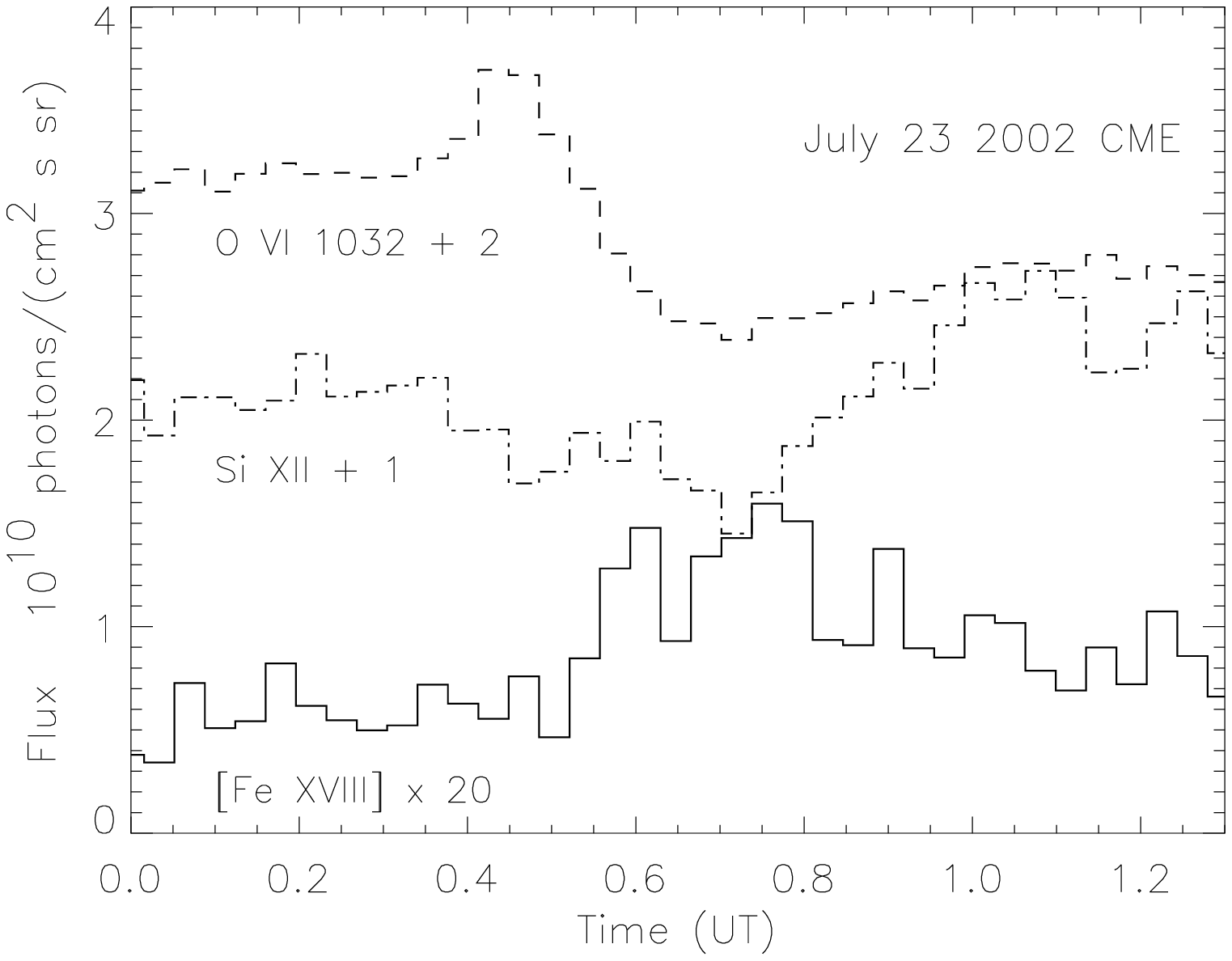}
\plotone{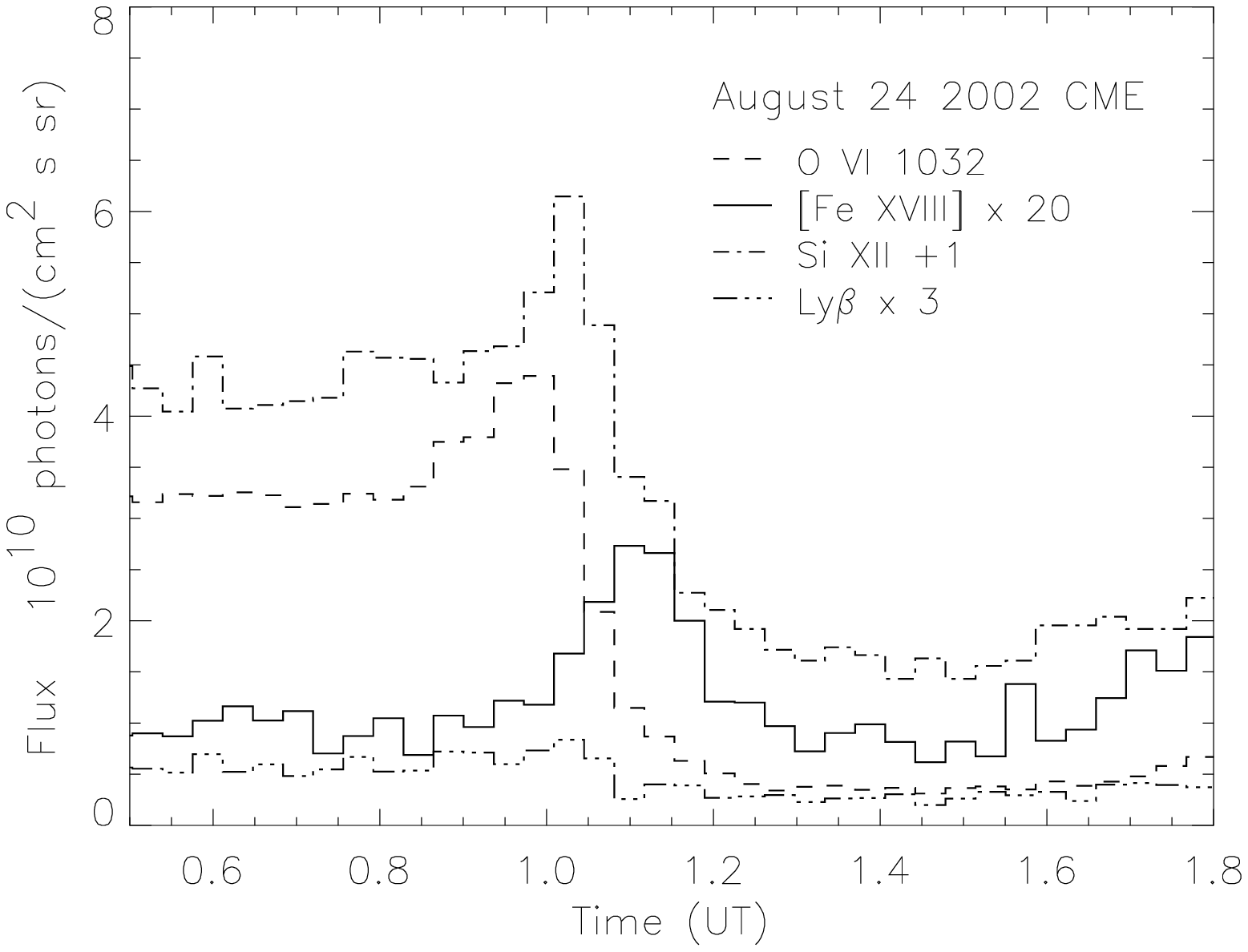}
\caption{ Figure 4a.  Evolution of the intensities of several spectral lines with time extracted
from a small (210\arcsec ) section of the UVCS slit centered on the bright transient [Fe XVIII]
feature for the 21 April event. The [Fe XVIII] and Ly$\beta$ intensities are scaled by factors
of 20 and 3 to be visible on the same plot.  b) Same for a 840\arcsec\/ section in the July event.
The O VI and Si XII lines have been shifted upwards to separate the lines for legibility.
c) Same for a 420\arcsec\/ section in the August event. The Si XII line has been shifted 
upwards to separate the lines for legibility.}
\end{figure}

\clearpage

\begin{figure}
\plottwo{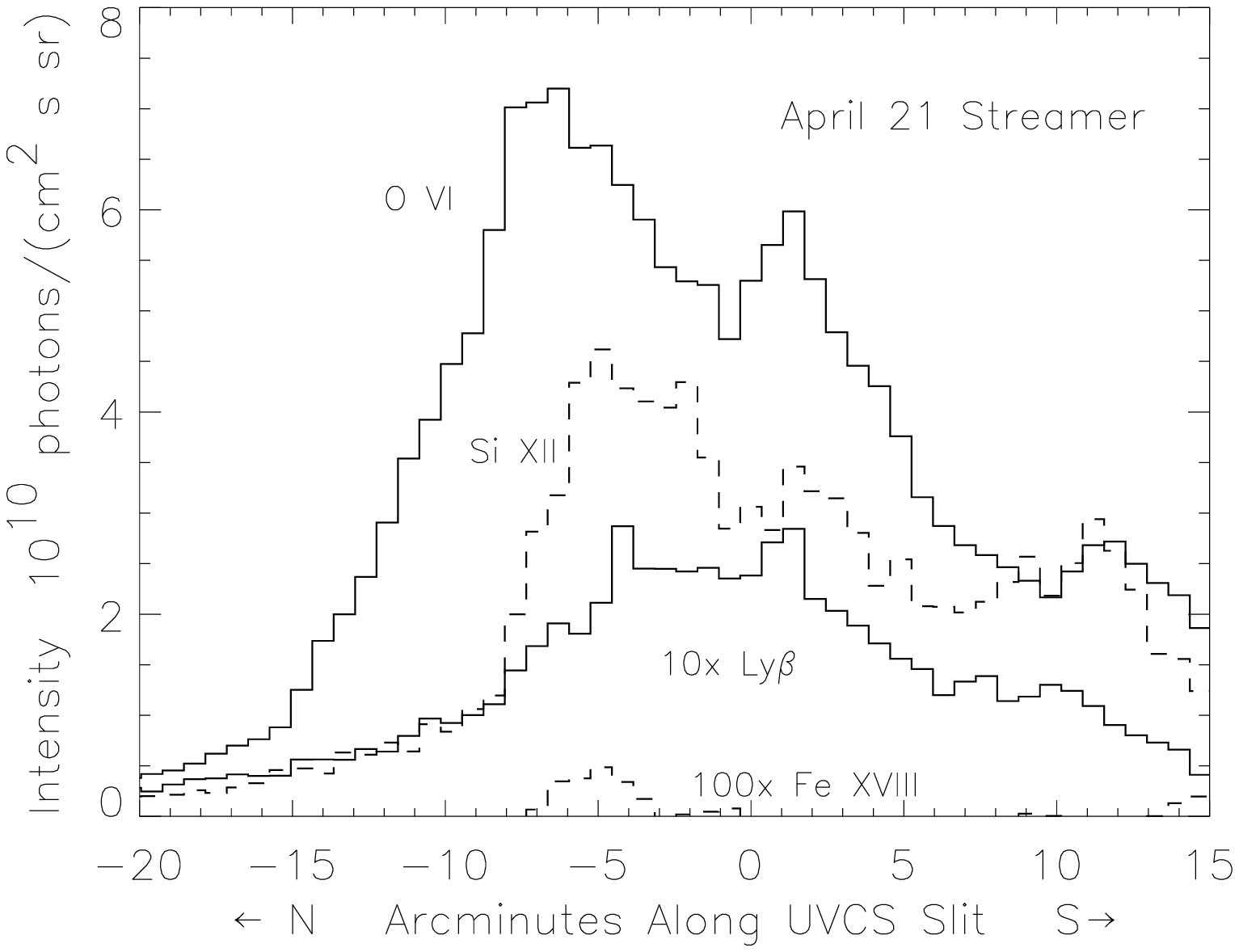}{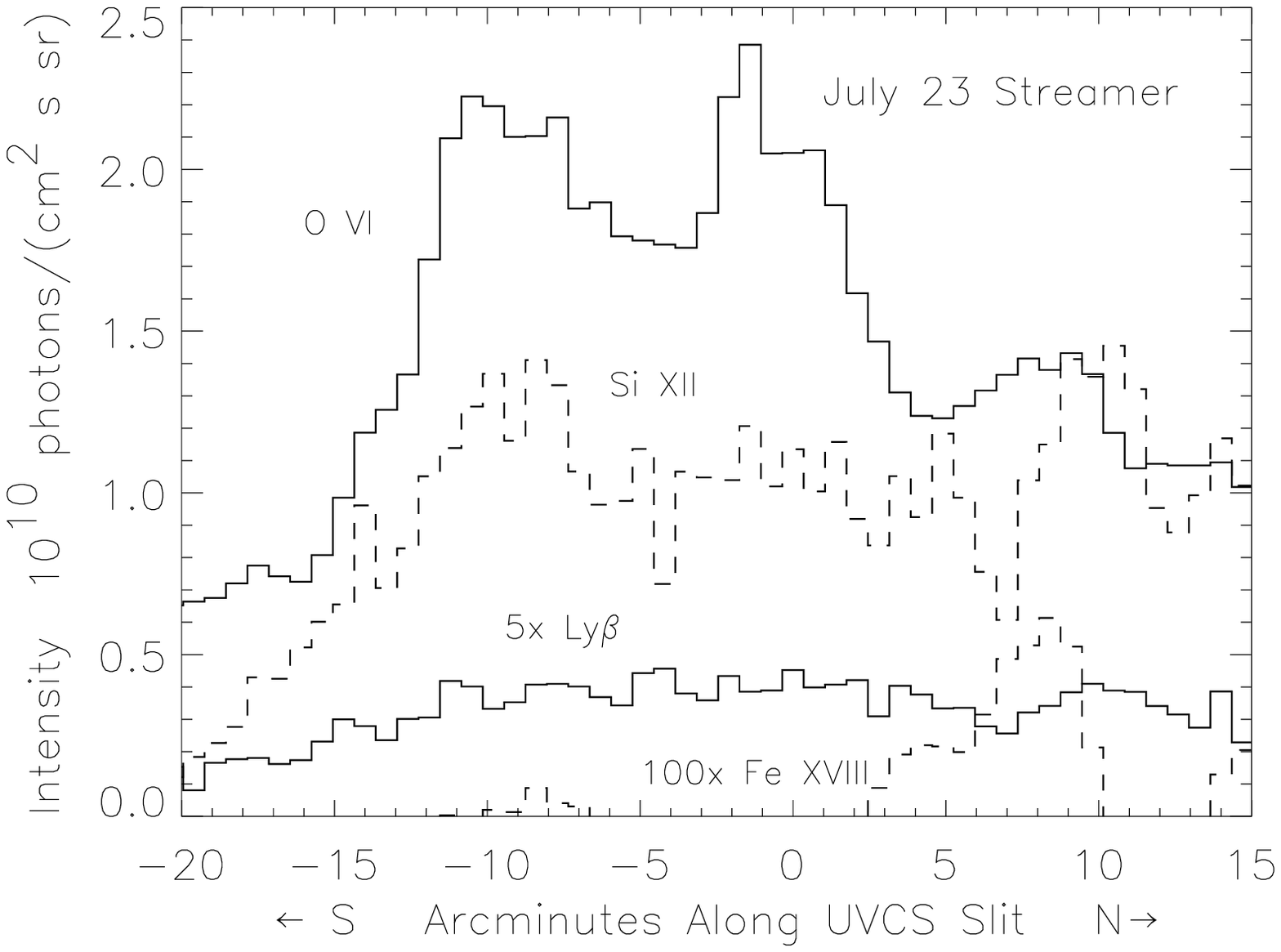}
\plotone{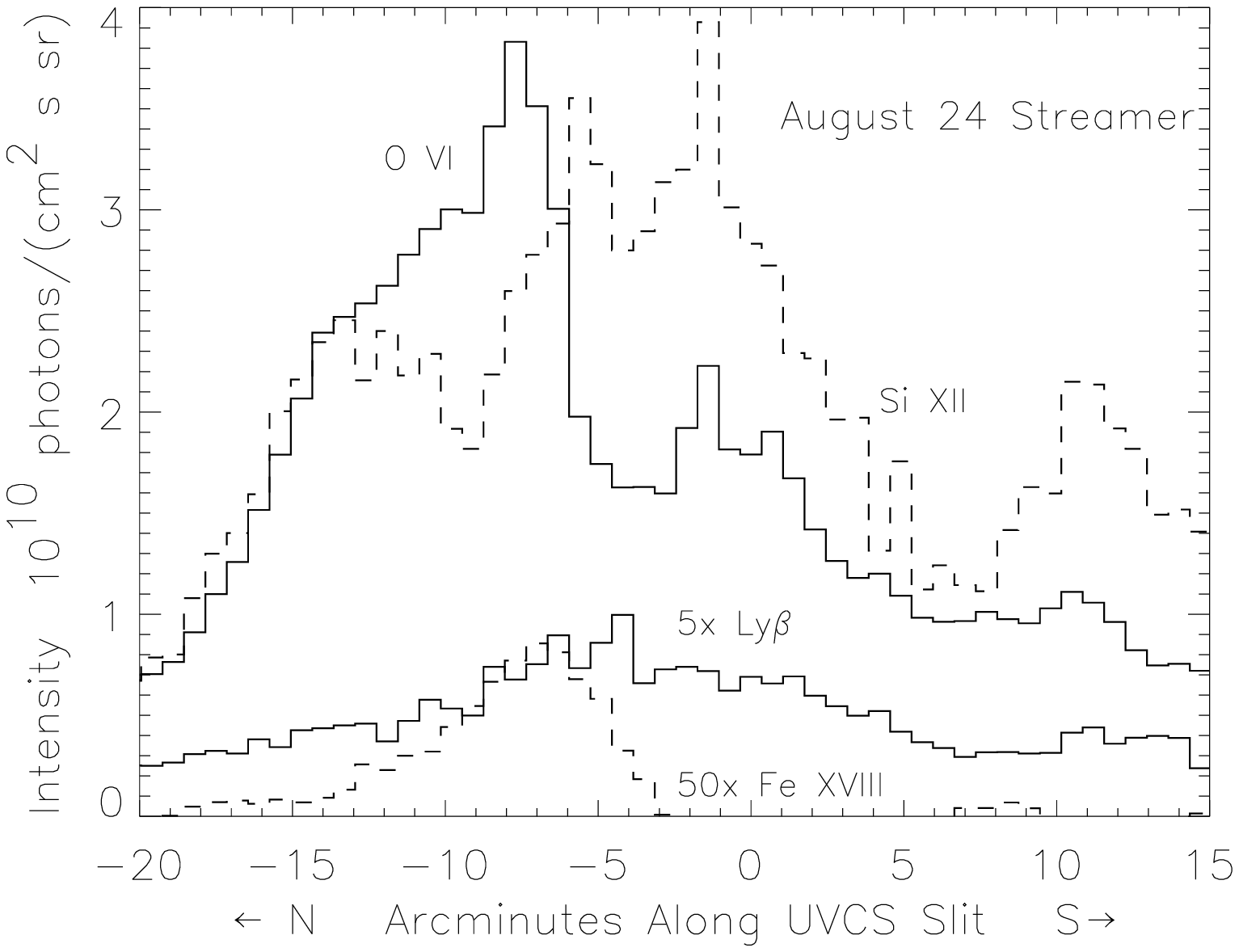}
\caption{ Figure 5a.  Intensities of the Ly$\beta$, O VI $\lambda$1032, Si XII and [Fe XVIII] lines along the
UVCS slit shortly before the 21 April event.  b) Same for 23 July event.
c) Same for the 24 August event.}
\end{figure}

\clearpage

\begin{figure}
\plotone{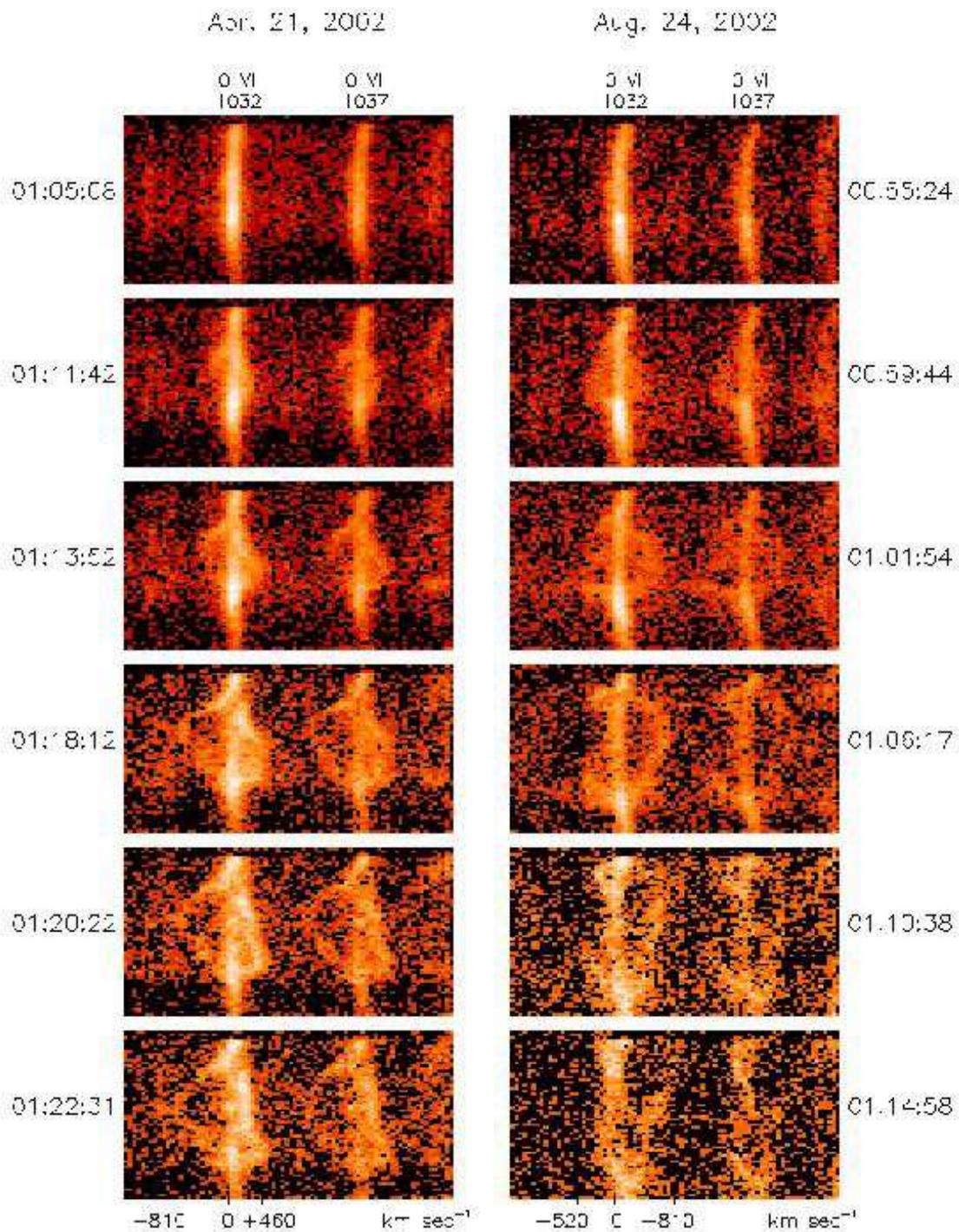}
\caption{ Figure 6.  Six longslit spectra for the April and August events showing the
O VI $\lambda \lambda$1032, 1038 lines, with Ly$\beta$ and the second order Si XII
$\lambda$521 lines faintly visible.  Velocities corresponding to Doppler shifts
of O VI $\lambda$1032 are shown on the horizontal scales, and UT times for each
exposure are shown to the left and right.}
\end{figure}
 
\clearpage

\begin{figure}
\plotone{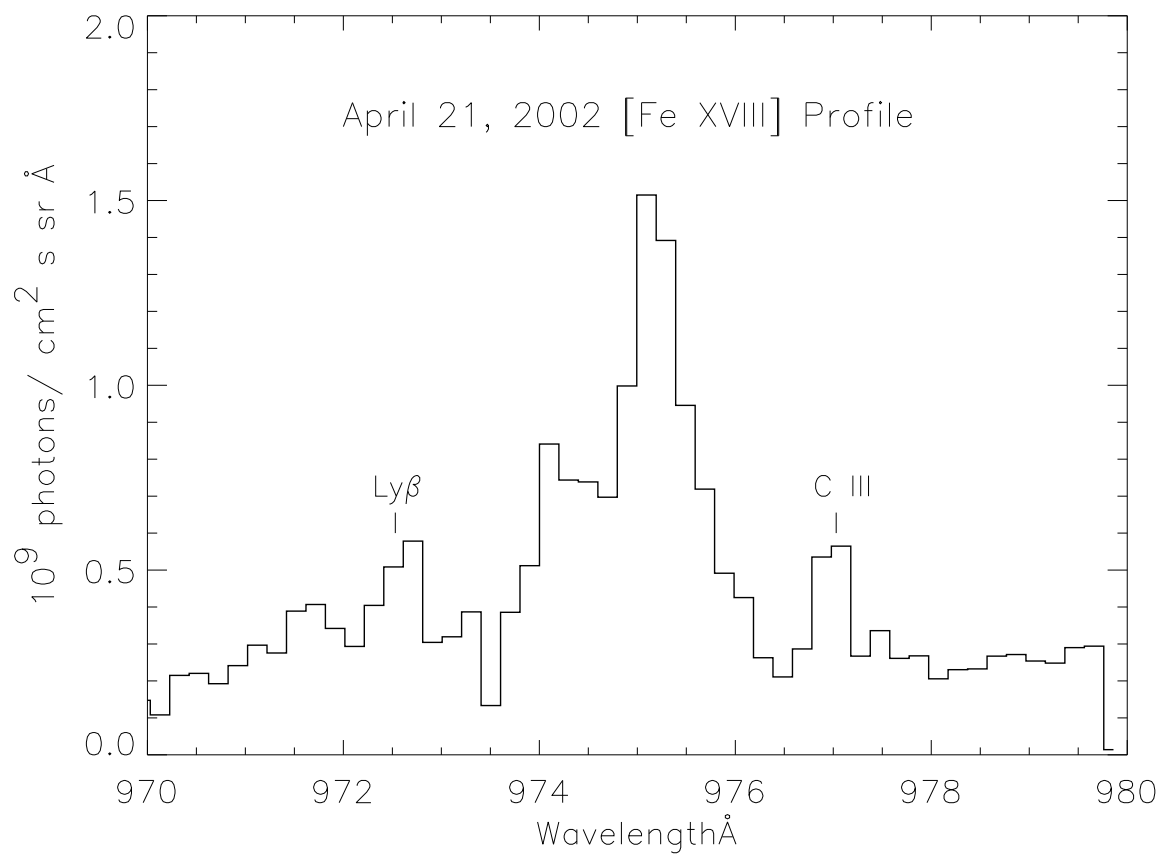}
\caption{Figure 7.  [Fe XVIII] line profile for the April event extracted from the sum
of 4 exposures (01:16 to 01:24 UT) over the -189\arcsec\/ to +21\arcsec\/ portion
of the slit.}
\end{figure}


\begin{thebibliography}{}

 
 
 
\bibitem[Amari et al. 2003]{amari}Amari, T., Luciani, J.F., Aly, J.J., Mikic, Z., \& Linker, J. 2003, ApJ, 585, 1073

\bibitem[Antiochos, Devore \& Klimchuk (1999)]{antiochos}Antiochos, S.K., Devore, C.R., and Klimchuk, J.A. 1999, ApJ, 512, 985
 

\bibitem[Ciaravella et al. 1997]{ciaravella97} Ciaravella, et al. 1997, ApJL, 491, 59

\bibitem[Ciaravella et al. 2000]{ciaravella00} Ciaravella, A., et al. 2000, ApJ, 529, 575

\bibitem[Ciaravella et al. 2002]{ciaravella02}Ciaravella, A., Raymond, J.C., Li, J., Reiser, P.,
Gardner, L.D., Ko, Y.-K., \& Fineschi, S.  2002, ApJ, 575, 1116

\bibitem[Dere et al. 2001]{dere}Dere, K.P., Landi, E., Young, P.R. and Del Zanna, G. 2001, ApJS, 134, 331


\bibitem[Endeve, Leer \& Holzer 2003]{endeve} Endeve, E., Leer, E., \& Holzer, T.E. 2003, preprint

\bibitem[Feldman et al. 1998]{feldman} Feldman, U. Sch\"{u}le, U., Widning, K.G., \& Laming, J.M. 1998, ApJ, 505, 999

\bibitem[Fontenla \& Poland]{fontenla} Fontenla, J.M., \& Poland, A.I. 1989, Sol. Phys. 123, 143

\bibitem[Gallagher et al. 2002]{gallaher2} Gallagher, P.T., Dennis, B.R., Krucker, S., Schwartz, R.A., \&
 Tolbert, A.K. 2002, Sol. Phys., 210, 341

\bibitem[Gallagher et al. 2003]{gallagher3} Gallagher, P.T., Lawrence, G.R., \& Dennis, B.R 2003, ApJL, 588, L53

\bibitem[Gardner et al. 2002]{gardner2002} Gardner, L.D., et al. 2002, in {\it The Radiometric Calibration of SOHO},
A. Pauluhn, M.C.E. Huber \& R. von Steiger, eds. (Noordwijk: ESA), p. 161

\bibitem[Gibson \& Low 1998]{gibsonlow}Gibson, S.E., and Low, B.C. 1998, ApJ, 493, 460

\bibitem[Hundhausen 1997]{hundhausen}Hundhausen, A.J. 1997, in {\it Coronal Mass Ejections}, N. Crooker,
J.A. Jocelyn \& J. Feynman, eds. (Washington: AGU), p. 1

\bibitem[Innes et al., 2001]{innes} Innes, D.E., Curdt, W., Schwenn, R., Solanki, S., Stenborg, G., \& 
McKenzie, D.E. 2001, ApJL, 549, 249

\bibitem[Ko et al. 2002]{koa}Ko, Y.-K., Raymond, J.C., Li, J., Ciaravella, A., Michels, J., Fineschi,
S., \& Wu, R. 2002, ApJ, 578, 979

\bibitem[Ko et al. 2003]{kob}Ko, Y.-K., Raymond, J.C., Lin, J., Lawrence, G., Li, J., \& Fludra, A. 2003,
ApJ, in press

\bibitem[Kohl \& Withbroe 1982]{kohlwith}Kohl, J.L., \& Withbroe, G.L. 1982, ApJ, 256, 263

\bibitem[Kohl et al. 1995]{kohl95} Kohl, J.L., et al. 1995, Solar Phys., 162, 313

\bibitem[Kohl et al. (1997)]{kohl97} Kohl, J.L., et al. 1997, Solar Phys., 175, 613

\bibitem[Kontar et al. 2003]{kontar} Kontar, E.P., Brown, J.C., Emslie, A.G., Schwartz, R.A., 
Smith, D.M., \& Alexander, R.C. 2003, preprint

\bibitem[Lepri et al., 2001]{lepri}Lepri, S.T., Zurbuchen, T.H., Fisk, L.A., 
Richardson, L.G., Cane, H.V., \& Gloeckler, G., 2001, JGR, 106, 29231

\bibitem[Li et al. 1998]{li} Li, J., Raymond, J.C., Acton, L.W., Kohl, J.L., Romoli, M., 
Noci, G., \& Naletto, G. 1998, ApJ, 506, 431

\bibitem[Lin 2002]{lin02} Lin, J. 2002, ChJAA, 2,539

\bibitem[Lin \& Forbes 2000]{linforbes}Lin, J., and Forbes, T.R. 2000, JGR, 105, 2375

\bibitem[Lin, Raymond \& van Ballegooijen]{lin03} Lin, J., Raymond, J.C., \& van Ballegooijen, A. 2003, ApJ, submitted

\bibitem[Low 2001]{low} Low, B.C. 2001, JGR, 106, 25141

\bibitem[Low \& Zhang (2002)]{lowzhang}Low, B.C., \& Zhang, M. 2002, ApJL 564, L53

\bibitem[Manchester 2001]{manchester}Manchester, W. IV 2001 ,ApJ, 547, 503

\bibitem[]{mancuso} Mancuso, S., Raymond, J.C., Kohl, J., Ko, Y.-K., Uzzo, M., \& Wu, R. 2002, A\& A, 383, 267 

\bibitem[Mazzotta et al. 1998]{mazzotta} Mazzotta, P., Mazzitelli, G., Colafrancesco, S., \& Vittorio, N.
1998, A\& AS, 133, 403

\bibitem[McMullin et al. 2002]{mcmullin}McMullin, D.R., Judge, D.l., Phillips, E.,
Moebius, E., Bochsler, P., Wurz, P., Hillenbach, M., \& Ipavich, F. 2002, in SOHO 11 Symposium,
ESA-SP -508

\bibitem[Moon et al. 2002]{moon} Moon, Y.-J., Choe, G.S., Wang, H., Park, Y.D., Gopalswamy, N., 
Yang, G., \& Yashiro, S. 2002, ApJ, 581, 694

\bibitem[Noci, Kohl \& Withbroe 1987]{noci}Noci, G., Kohl, J.L. \& Withbroe, G.L. 1987, ApJ, 315, 706

\bibitem[Parenti et al. 2000]{parenti} Parenti, S., Bromage, B.J., Poletto, G., Noci, G., 
Raymond, J.C., and Bromage, G.E.  2000, A\& A, 363, 800

\bibitem[Pike \& Mason 2002]{pike} Pike, C.D., and Mason, H.E. 2002, Sol. Phys., 206, 359

\bibitem[Raymond 2002]{raymond02} Raymond, J.C. 2002, in {From Solar Min to Max: Half a Solar Cycle with SOHO},
A. Wilson, ed., (Noordwijk: ESA), p. 421

\bibitem[Raymond et al. 1997]{raymond97} Raymond, J.C., et al. 1997, Sol. Phys. 197, 645

\bibitem[]{raymond00} Raymond, J.C., Thompson, B.J., St. Cyr, O.C., Gopalswamy, N., Kahler, S., Kaiser, M., Lara, A.,
Ciaravella, A., Romoli, M., \& R. O'Neal, R., 2000, GRL, 27, 1493

\bibitem[]{roussev} Roussev, I.I., Forbes, T.G., Gombosi, T.I., Sokolov, I.V., DeZeeuw, D.L., \& Birn, J. 2003, ApJL, 588, L45

\bibitem[Rugge \& McKenzie 1985]{rugge} Rugge, H.R., \& McKenzie, D.L. 1985, ApJ, 297, 338

\bibitem[Schmahl \& Hildner 1977]{schmahl} Schmahl, E.J., \& Hildner, E. 1977, Sol. Phys. 55, 473

\bibitem[Sch\"{u}le et al. 2000]{schule} Sch\"{u}le, U., Wilhelm, K., Hollandt, J., Lemaire, P.,
\& Pauluhn, A. 2000, A\& A, 354, L71

\bibitem[Share et al. (2003)]{share}Share, G.H., Murphy, R.J., Dennis, B.R., Schwartz, R.A.,
Tolbert, A.K., Lin, R.P., \& Smith, D.M. 2002, Solar Physics, 210, 357

\bibitem[Share et al. (2003)]{share2}Share, G.H., Murphy, R.J., Skibo, J.G., Smith, D.M., Hudson, H.S., Lin, R.P.,
Shih, A.Y., Dennis, B.R., Schwartz, R.A. \& Kozlovsky, B. 2003, preprint

\bibitem[St. Cyr et al. 2000]{stcyr} St.Cyr, O.C., Burkepile, J.T., Hundhausen, A.J., \& Lecinski, A.R. 2000, JGR, 104, 12493

\bibitem[Strachan et al. 2002]{strachan} Strachan, L., Suleiman, R., Panasyuk, A.V., Biesecker, D.A.,
\& Kohl, J.L. 2002, ApJ, 571, 1008

\bibitem[Suess \& Nerney 2002]{suess} Suess, S.T. \& Nerney, S.F. 2002, ApJ, 565, 1275

\bibitem[Uzzo et al. 2003]{uzzo} Uzzo, M., Ko, Y.-K., Raymond, J.C., Wurz, P., \&
Ipavich, F.M. 2003, ApJ, 585, 1062
\bibitem[Vernazza \& Reeves 1978]{vr} Vernazza, J.E., \& Reeves, E.M. 1978, ApJS, 37, 485.


\bibitem[Wang et al. (2002)]{wang}Wang, T.J., Solanki, S.K., Innes, D.E., \& Curdt, W. 2002,
preprint

\bibitem[Webb et al. 2003]{webb} Webb, D.F., Burkepile, J., Forbes, T.G., \& Riley, P. 2003, preprint

\bibitem[Woods et al. 2000 ]{woods} Woods, T.N., Tobiska, W.K., Rottman, G.J., \& Worden, J.R. 2000,
J. Geophys. Res., 105, 27195

\bibitem[Young et al. 2003]{young} Young, P.R., Del Zanna, G., Landi, E., Mason, H.E., \& Landini, M.
2003, ApJS, 144, 135

\end{thebibliography}
\end{document}